\documentclass[12pt]{article}
\pdfoutput=1

\usepackage[bulletsep]{collref}
\usepackage{amssymb,graphicx}
\usepackage[intlimits]{amsmath}
\usepackage{bbm}
\usepackage[small]{subfigure}
\newcommand{\rf}[1]{(\ref{#1})}

\usepackage{MnSymbol}

\makeatletter \@addtoreset{equation}{section} \makeatother

\makeatletter
\let\old@startsection=\@startsection
\let\oldl@section=\l@section
\renewcommand{\@startsection}[6]{\old@startsection{#1}{#2}{#3}{#4}{#5}{#6\mathversion{bold}}}
\renewcommand{\l@section}[2]{\oldl@section{\mathversion{bold}#1}{#2}}
\makeatother

\makeatletter
\let\old@makecaption=\@makecaption
\def\@makecaption{\small\old@makecaption}
\makeatother

\renewcommand{\geq}{\geqslant}

\begin{document}


\thispagestyle{empty}
\today
\begin{flushright}\footnotesize
\texttt{Imperial-TP-AT-2018-03}\\
\texttt{NORDITA-2018-033} \\
\texttt{UUITP-15/18}

\end{flushright}

\renewcommand{\thefootnote}{\fnsymbol{footnote}}
\setcounter{footnote}{0}

\begin{center}
{\Large\textbf{\mathversion{bold} 
Precision   matching
 of circular Wilson  loops\\   and
  strings  in $AdS_5 \times S^5$     
}
\par}

\vspace{0.3cm}

\textrm{Daniel~Medina-Rincon$^{1,2}$, Arkady A. Tseytlin$^{3}$\footnote{Also at Lebedev Institute, Moscow, Russia}\\ and
Konstantin~Zarembo$^{1,2,4}$\footnote{Also at ITEP, Moscow, Russia}}
\vspace{4mm}

{\small 
\textit{${}^1$Nordita, KTH Royal Institute of Technology and Stockholm University\\
Roslagstullsbacken 23, SE-106 91 Stockholm, Sweden}\\
\textit{${}^2$Department of Physics and Astronomy, Uppsala University\\
SE-751 08 Uppsala, Sweden}\\
\textit{${}^3$Blackett Laboratory, Imperial College, London SW7 2AZ, UK}\\
\textit{${}^4$Hamilton Mathematical Institute, Trinity College, Dublin 2, Ireland}\\
\vspace{0.2cm}
\texttt{d.r.medinarincon@nordita.org, tseytlin@imperial.ac.uk, zarembo@nordita.org}
}



\par\vspace{0.5cm}

\textbf{Abstract} \vspace{3mm}

\begin{minipage}{13cm}
Previous   attempts  to match the exact ${\cal N}=4 $  super Yang-Mills   expression  for the expectation value of the 
$1\over 2$-BPS circular Wilson loop  with the semiclassical $AdS_5 \times S^5$     string theory  prediction 
were not successful  at the first   subleading order. 
 There was a   missing  prefactor  $\sim  \lambda^{-3/4}$  which  could be attributed to 
 the  unknown normalization of the string path integral measure. 
Here  we  resolve this problem  by computing the ratio  of the 
string partition functions 
 corresponding to the circular Wilson  loop  and the  special 
 $1\over 4$-supersymmetric  latitude  Wilson  loop.
 The fact that the  latter    has a trivial expectation value  in the gauge theory  allows  us to relate 
 the prefactor to the contribution of the  three zero modes  of the ``transverse"   fluctuation operator  in 
 the 5-sphere directions. 
 
\end{minipage}

\end{center}

\vspace{1.5cm}


\def \la{\label}
 \def \ha {\te {1\ov 2}}
\def \te {\textstyle}\def \bi {\bibitem} 
\def \ve {\Lambda}
\def \s {\sigma} \def \ov {\over} 
\def \ci {\cite} \def \td {\tilde} 
  \def \a {\alpha}
\def \iffa {\iffalse} \def \N {{\cal N}} 
\def \foot {\footnote}
\def \ed {\end{document}}
\def \be {\begin{equation}} \def \OO {{\cal O}} 
\def \ee {\end{equation}}  \def \CC  {{\rm C}} \def \sql {\sqrt \lambda} \def \nn {{n}}\def \LL {{\rm L}} \def \l  {\lambda}
\setcounter{page}{1}
\renewcommand{\thefootnote}{\arabic{footnote}}
\setcounter{footnote}{0}

\newpage

\tableofcontents

\

\section{Introduction}

One of the   central elements of the   gauge-string correspondence  is that the expectation   value of the Wilson loop 
operator should  be given by the string path integral  \ci{Polyakov:1997tj}. 
In the context of the maximally supersymmetric AdS/CFT  duality this translates, in particular, 
into the relation   between the  large $N$ expectation value  of the locally-supersymmetric Wilson loop
in $\N=4$  SYM  theory  \ci{Maldacena:1998im,Rey:1998ik}
and the $AdS_5 \times S^5  $ superstring  path integral      on a disc   with   appropriate boundary conditions
on a contour at the boundary of  $AdS_5$ and  in $S^5$. 
Checking this  precisely  is  non-trivial  
as it  requires a  careful normalization of the string path integral  (which is subtle  even  in flat  space, cf. \ci{Alvarez:1982zi}). 

The  simplest  well-defined  example is the   $\frac{1}{2}$-BPS    circular WL \ci{Berenstein:1998ij},\ci{Drukker:1999zq} 
 the  expectation  value  of which can be   found  exactly  on the gauge theory side 
 due to underlying superconformal symmetry  \cite{Erickson:2000af,Drukker:2000rr,Pestun:2007rz}.  Considering the planar limit
 and expanding at strong  't Hooft coupling $\lambda =g_{\rm YM}^2N_c$,  
 one finds
 \be
\la{1}
 W_{\CC}  = \frac{2}{\sqrt\lambda}\,I_{1}(\sqrt\lambda) = 
{\frac{\sqrt 2}{\sqrt \pi}}\, { {1\ov ( \sqrt \lambda)^{3/2} } }\, e^{\sqrt\lambda}\,\Big(1-\frac{3}{8\,\sqrt\lambda}+\cdots\Big),  \qquad \lambda \gg 1\ .
\ee
The   corresponding    $AdS_5 \times S^5  $   superstring  partition function expanded near  the 
$AdS_2$   minimal surface ending  on a circle at the boundary 
of $AdS_5$   should   have 
  the following expansion in  inverse string tension $T={\sql \ov 2 \pi}$ \ci{Drukker:2000ep}\foot{To simplify  the notation we will not use  separate   letters   for the gauge theory and string theory   results as they should be eventually  shown to be equal.} 
\be \la{2} 
W_\CC  = \nn_0\,   \exp \Big[ { \sqrt \lambda  + c_1  + {c_2 \ov \sqrt \lambda} + \OO (\l^{-1/2}) } \Big]
\ , \ee  
where $\sqrt \lambda$  term  comes from the classical string action,
 $c_1$ is  one-loop string sigma-model   correction, etc. Here 
$\nn_0$ is  a potential  overall factor  of  normalization  of the string path integral measure on a disc. 

The computation  of the   superstring fluctuation  determinants in \cite{Kruczenski:2008zk,Kristjansen:2012nz}  and \ci{Buchbinder:2014nia}
gave  the  one-loop value $c_1= - \ha \log  (2 \pi) $,  reproducing   the $\pi$-factor in \rf{1},  with the remaining $\lambda^{-3/4}$ 
and power   of 2  factor   to  be attributed to  the presence of  the  $\nn_0$   normalization factor.  
It was  suggested in \ci{Drukker:2000rr} that $\lambda^{-3/4}$  may be related to the  subtraction of the 
  M\"obius   symmetry volume 
(or  the ghost determinant   zero modes) on the disc (see  also \ci{Ambjorn:2011wz}). 
 
 Our aim here will be to reproduce the exact prefactor  in \rf{1}  and thus establish  finally  the precise matching between 
 the gauge  theory  and string theory  predictions  at  the two leading orders  at strong coupling.
   Instead of  computing the normalization $\nn_0$ directly   we  will   consider
  the ratio   of the circular WL   expectation  value to the expectation value of 
  $\frac{1}{4}$-supersymmetric WL \ci{Zarembo:2002an}   generalizing circular loop
     to the case when the string surface ends also  on a big circle of $S^5$.    
   The   residual  global supersymmetry (this loop  preserves      2  out of 8 $Q$-supercharges) 
  implies  the trivial expectation   value  for  this loop    for all  values   of $\l$  
  \be  W_{\LL}  =1 \ , \la{3}\ee
  as anticipated in \cite{Zarembo:2002an}. This non-renormalization theorem can be rigorously proven using superspace arguments in the $\mathcal{N}=4$ gauge theory \cite{Guralnik:2003di,Guralnik:2004yc}.
  While  checking \rf{3}     directly on the string side would  again  require the knowledge of  the  string measure factor
  $n_0$,    one   may  consider the ratio of the   two expectation values 
  $ W_{\CC}/W_{\LL} $   in which the nuisance factor $n_0$ cancels out.
 Assuming non-renormalization of $W_{\rm L}$,  
the  1-loop 
  string theory computation should  reproduce  the precise   factor  appearing in \rf{1}. 
  The appearance of  $\lambda^{-3/4}$   can then be attributed to  the normalization of the three   zero modes  of the $S^5$
   fluctuation operator that arise due to degeneracies of the minimal surface for $W_{\rm L}$ \ci{Zarembo:2002an}\foot{The  absence of the   $\lambda^{-3/4}$    factor  in \rf{3} may be then 
    understood as a result of the  cancellation between  the   normalizations   of the 
     3   M\"obius symmetry zero modes of the ghost operator in conformal gauge
      and the 3 zero modes of the $S^5$  fluctuation operator.}.
       
       We used the label  L in \rf{3} to indicate that this  $\frac{1}{4}$-supersymmetric 
        WL      is a special  case  ($\theta_0={\pi\ov 2}$)   of  a  more general 
    $\frac{1}{4}$-BPS    ``latitude"  WL  \cite{Drukker:2005cu,Drukker:2006ga}  depending on the  latitude angle parameter $\theta_0$: 
    \be\la{4}    W_{\LL} \equiv W_{\LL} (\l, \theta_0={\pi\ov 2})   \ . \ee
    The case of  $\theta_0=0$  corresponds to the $\frac{1}{2}$-BPS  circular WL    and in general 
   the exact   expression  for  the expectation value of this loop $ W_{\LL} (\theta_0)$ 
is  given by    \rf{1}   with the replacement $\l \to \l \cos^2 \theta_0$, i.e. $W_{\LL} (\l, \theta_0)= W_\CC (\l \cos^2 \theta_0)$ 
 \ci{Drukker:2006ga,Pestun:2009nn}.
 The   computation of    the ratio $W_{\LL} (\l, \theta_0)/W_\CC$ on the string side was        suggested  in 
   \cite{Forini:2015bgo}  and developed in \ci{Faraggi:2016ekd}  and  later led to  precise checks  of the correspondence   with gauge theory result,  
   first  in small $\theta_0$  expansion 
   \ci{Forini:2017whz}  and then for  general $0 < \theta_0 < {\pi\ov2}$  \ci{Cagnazzo:2017sny},  reproducing 
   \be \la{5} 
   {W_\LL (\l, \theta_0)  \ov W_\CC(\l) } = 
   \exp \Big[ { -\sql (1 - \cos \theta_0)   - {3\ov 2} \ln \cos \theta_0 + \OO (\l^{-1/2}) }\Big]  \ , \ \ \ \ \ \ \     \theta_0 < {\pi\ov2} \ .      \ee
   The  aim of the present paper is to extend the   discussion in  \ci{Cagnazzo:2017sny} to 
   the special case of $\theta_0={\pi\ov 2}$  when  the $S^5$ fluctuation operator develops   3  zero   modes
   and to show  that   in this case $ W_{\CC}/W_{\LL} $ is also in full agreement  with \rf{1}, 
   including  the enigmatic  $\l^{-3/4}$  and $\sqrt 2$ factors.

We shall start in section 2   with a brief review  of the  definitions of $W_\CC$   and $W_\LL$   in gauge theory 
 and  the  corresponding  minimal string surfaces in $AdS_5 \times S^5$.  
 Semiclassical  expansion near a  minimal string surface   will be  discussed in section 3, paying 
 special attention to the contribution of zero modes of the quadratic  fluctuation operator present in the  special  
$\theta_0={\pi\ov 2}$  latitude case.  The relevant  determinant will be computed  in section 4 demonstrating that the string theory 
 prediction  for  the ratio ${W_\LL / W_\CC } $   is in perfect  agreement with the gauge-theory results  \rf{1},\rf{3}
 expanded   at large $\l$. Properties of the conformal anomaly in the presence of zero modes, used in the intermediate steps of the derivation, are reviewed in the Appendix.

\section{Setup}

\subsection{Wilson loops}

The Wilson loop expectation value in $\mathcal{N}=4$ super-Yang-Mills theory  \cite{Maldacena:1998im} is given by
\begin{equation}\la{21}
 W(C;\mathbf{n})=\frac{1}{N}
 \Big \langle 
 \mathop{\mathrm{tr}}{\rm P}\exp\Big[i\int_{C}^{}d\tau \,
 \big(\dot{x}^\mu A_\mu +i|\dot{x}|n^I\Phi _I\big)
 \Big]
 \Big\rangle.
\end{equation}
The spacial contour will always be the unit circle: $x^\mu =(\cos\tau,\sin\tau ,0,0 )$, while for the scalar coupling we consider two different cases: the pure circular loop with constant coupling to scalars and the 
 latitude. The coupling of the Wilson loop to scalars is parameterized by a unit 6-vector $n^I$ on $S^5$, best represented for our purposes as an $S^1\times S^3$ fibration over an interval,  
\begin{equation}\label{n-S5}
 \mathbf{n}=\left(\cos\theta \,\mathbf{k}, \ \sin\theta \,\cos\varphi ,\ \sin\theta \,\sin\varphi \right),
\end{equation}
where $\mathbf{k}$ is a  unit 4-vector. The circular loop and the special latitude correspond to
\begin{eqnarray}
 {\rm C}:&& \theta =0,\qquad \mathbf{k}=\left(1,0,0,0\right), \qquad \varphi ={\rm any},
\nonumber \\
{\rm L}:&& \theta =\frac{\pi }{2}\,,\qquad \mathbf{k}={\rm any}, \qquad \varphi =\tau .
\end{eqnarray}
On the north pole of the 5-sphere $S^1$ shrinks to zero size and consequently $\varphi $ can take any value, while on the equator $S^3$ shrinks to zero size and $\mathbf{k}$ can be arbitrary. The vector $\mathbf{k}$ is not an extra parameter of the contour but will reappear as such in the string-theory calculation where the string will move away from the locus where $S^3$ is shrunk to a point. 

As discussed above,  the 
 expectation value of the ${1\ov 4}$-supersymmetric  latitude is trivial, and its ratio to the circular loop can be calculated exactly
 (cf. \rf{1},\rf{3}) 
\begin{equation}\label{prediction}
 \frac{W_{\rm L}}{W_{\rm C}}=\frac{\sqrt{\lambda }}{2I_1\left(\sqrt{\lambda }\right)}\stackrel{\lambda \rightarrow \infty }{\simeq }
 \sqrt{\frac{\pi }{2}}\,\lambda ^{\frac{3}{4}}\,{\rm e}\,^{-\sqrt{\lambda }}.
\end{equation}

\subsection{String solutions}

In string theory, the Wilson loop is represented by a disc partition function which at strong coupling is saturated by the area law in $AdS_5\times S^5$:
\begin{equation}\label{semicl}
 W(C;\mathbf{n})\stackrel{\lambda\rightarrow \infty  }{=}
 n_0\mathop{\mathrm{Sdet}^{-\frac{1}{2}}\mathbbm{K}}\,{\rm e}\,^{-\frac{\sqrt{\lambda }}{2\pi }\,A_{\rm min}},
\end{equation}
where $A_{\rm min}$ is the (regularized) minimal area, that depends on the contour at hand, $\mathbbm{K}$ is the quadratic form of string fluctuations around the minimal surface and $n_0$ is the measure factor discussed in the introduction.\footnote{This semiclassical formula gets modified if the minimal surface is degenerate and $\mathbbm{K}$ develops zero modes. Modifications are non-trivial and discussed in detail later.}

The metric of  $AdS_{5}\times S^5$ that corresponds to the Poincar\'e  coordinates  in $AdS_5$ 
 and to the  parameterization  (\ref{n-S5}) of the sphere is
\begin{equation}
 ds^2 = \frac{dX_\mu ^2+dZ^2}{Z^2}
 +d\theta ^2+\sin^2\theta \,d\varphi ^2+\cos^2\theta \,d\Omega _{S^3}^2 \ . 
\end{equation}
The AdS part of the string solution is the same for the circle and latitude: 
\begin{equation}
{\rm C/L}: \qquad X^\mu =\frac{\left(\cos\tau ,\sin\tau ,0,0\right)}{\cosh\sigma }\,,\qquad 
 Z=\tanh\sigma  \ .  
\end{equation}
For the circle the string always stays at $\theta =0$ and the dynamics on $S^5$ are trivial, while for the latitude the string has a non-trivial profile in $S^5$ \ci{Zarembo:2002an}:
\begin{equation}\label{L-S5}
{\rm L}: \qquad   \cos\theta =\tanh\sigma ,\qquad \varphi =\tau\ , 
\end{equation}
where $ \sigma \in (0,  \infty)$ and $\tau \in (0, 2\pi)$ are   the world-sheet coordinates. 
The unit 4-vector $\mathbf{k}$ remains arbitrary, meaning that there is a whole three-parametric family of solutions. 
All these solutions are different, because $\theta <\pi /2$ in the interior of the worldsheet and 
consequently $S^3$ inflates to a finite size away from the boundary at $\s=0$ ($Z=0$). 

Both of these solutions are in the conformal gauge, i.e.  the induced world-sheet metric is 
\begin{equation}\label{g-induced}
 ds_{\rm ind}^2=\Omega ^2\left(d\tau ^2+d\sigma ^2\right),
\end{equation}
with the conformal factors  being 
\begin{equation}\label{confactors}
 \Omega _{\rm C}^2=\frac{1}{\sinh^2\sigma }\,,\qquad \qquad 
 \Omega _{\rm L}^2=\frac{1}{\sinh^2\sigma }+\frac{1}{\cosh^2\sigma }\,.
\end{equation}

\section{Semiclassical string quantization}

The semiclassical formula (\ref{semicl}) can be readily applied to the circle, but has to be modified for the latitude to take into account the three-parametric degeneracy in the minimal surface. We briefly review the general formalism of collective coordinates and then apply it to the case of the latitude Wilson loop.

Suppose that a path integral of some field theory is saturated by a classical field configuration $\Phi _{\rm cl}$ that depends on $n$ parameters $\varphi _i$.  These parameters are traded for collective coordinates by the change of variables in the path integral. The general formula for the semiclassical partition function is
\begin{equation}
 \int_{}^{}\mathcal{D}\Phi \,\,{\rm e}\,^{-\frac{1}{\hbar}\,S\left[\Phi \right]}
 \stackrel{\hbar\rightarrow 0}{\simeq} \int 
 \prod_{i=1}^{n}\frac{d\varphi _i}{\sqrt{2\pi \hbar}}\,\,
 \det_{ij}{}^{\frac{1}{2}}\left\langle \frac{\partial \Phi _{\rm cl}}{\partial \varphi _i}\,,\,\frac{\partial \Phi _{\rm cl}}{\partial \varphi _j}\right\rangle
 \mathop{\mathrm{Sdet}}\nolimits '{}^{-\frac{1}{2}}\mathbbm{K}
 \ {\rm e}\,^{-\frac{1}{\hbar}\,S\left[\Phi _{\rm cl}\right]}, \la{31}
\end{equation}
where $\mathbbm{K}$ is the quadratic form of the action expanded around the classical solution:
\begin{equation}
 \mathbbm{K}=\frac{\delta ^2S}{\delta \Phi \,\delta \Phi }\,.
\end{equation}
 The quadratic form has zero modes, because
\begin{equation}
 \mathbbm{K}\cdot \frac{\partial \Phi _{\rm cl}}{\partial \varphi _i}=
 \frac{\partial }{\partial \varphi _i}\,\,\frac{\delta S}{\delta \Phi }=0,
\end{equation}
and those should be omitted (which is marked by prime on $\mathop{\mathrm{Sdet}}$), as they are already accounted for by the collective coordinates. The first determinant factor in \rf{31}  is the Jacobian of this transformation. Finally, the factor of $1\ov \sqrt{2\pi \hbar}$ comes from the missing Gaussian integral over each of the $n$ zero modes.\foot{The  standard normalization of the measure  is such that a Gaussian integral 
over each quadratic fluctuation mode should not depend on $\hbar$.}

We can now apply this general result to the latitude normalized by the expectation value of the circle. According to the AdS/CFT dictionary, the $\hbar$ of string theory is the inverse string tension \be 
\hbar=\frac{2\pi }{\sqrt{\lambda }}\,.
\ee
From (\ref{n-S5}), (\ref{L-S5}) we find:
\begin{equation}
 \frac{\partial \mathbf{n}_{\rm cl}}{\partial \varphi _i}=\left(\tanh\sigma \,\frac{\partial \mathbf{k}}{\partial \varphi _i}\,,0,0\right),
\end{equation}
where $\varphi _i$ are the three angles on the three-sphere. Since no other factor depends on $\mathbf{k}$, the integral over the collective coordinates just gives the volume of $S^3$:
\begin{equation}
 \int_{S^3}^{}d^3\varphi \, \det_{ij}{}^{\frac{1}{2}}\frac{\partial \mathbf{k}}{\partial \varphi _i}\cdot \frac{\partial \mathbf{k}}{\partial \varphi _j}
 ={\rm Vol}(S^3)=2\pi ^2.
\end{equation}
Taking into account that  the (regularized)   minimal area $S_{\rm cl}=-2\pi $ for the circle \cite{Berenstein:1998ij,Drukker:1999zq} and $0$ for the latitude  \ci{Zarembo:2002an}\footnote{The supersymmetric latitude belongs to a more general class of BPS string solutions, owing their existence to a non-integrable almost complex structure on $AdS_5\times S^5$.  All of these surfaces can be shown to have zero regularized area \cite{Dymarsky:2006ve}.}, one finds:
\begin{equation}\label{WL/WC}
 \frac{W_\LL}{W_\CC}=\frac{2\pi ^2}{\left(2\pi \right)^3}\,\lambda ^{\frac{3}{4}}
 \,{\rm e}\,^{-\sqrt{\lambda }}\left\langle \psi _0\left|\Omega _\LL^2\right|\psi _0\right\rangle^{\frac{3}{2}}\,\frac{\mathop{\mathrm{Sdet}}{}^\frac{1}{2}\mathbbm{K}_\CC}{\mathop{\mathrm{Sdet}}'{}^\frac{1}{2}\mathbbm{K}_\LL},
\end{equation}
where
\begin{equation}\label{psi-noll}
 \psi _0=\tanh\sigma 
\end{equation}
is the zero mode of the fluctuation operator on $S^5$.

 The norm of the zero mode is defined with respect to the induced metric on the string worldsheet and contains the scale factor
 $\Omega$. We reserve the bracket notation for the conventional unit norm, thus
\begin{equation}\label{curved<>}
 \left\langle \psi _1,\psi _2\right\rangle=
 \int_{}^{}d\tau \,d\sigma \,\Omega ^2\psi _1^*\psi _2=
 \left\langle \psi _1\left|\Omega^2\right|\psi _2\right\rangle.
\end{equation}

The explicit form of the one-loop contribution has been worked out by applying the general formalism of \cite{Drukker:2000ep,Forini:2015mca} to the minimal surface of the circular loop \cite{Drukker:2000ep,Kruczenski:2008zk} and the latitude \cite{Forini:2015bgo,Faraggi:2016ekd}:
\begin{align}\label{Z-1-loop}
\mathop{\mathrm{Sdet}}\mathbbm{K} = \frac{{{{\det }^{3}}{\mathcal{K}_1}\, {{\det }^{3}}{\mathcal{K}_2}\, {{\det }}{\mathcal{K}_{3 + }}\, {{\det }}{\mathcal{K}_{3 - }}}}{{{{\det }^4}{\mathcal{D}_ + }\,{{\det }^4}{\mathcal{D}_ - }}}\,,
\end{align}
where $\mathcal{K}_1$ corresponds to three fluctuation modes on $AdS_5$, $\mathcal{K}_2$ describes three ``transverse" modes on $S^5$, while the remaining two $S^5$ modes mix and result in $\mathcal{K}_{3\pm}$. 
The Dirac operators $\mathcal{D}_{\pm}$ originate from expanding the Green-Schwarz action on $AdS_5\times S^5$ to quadratic order in fermions in a particular kappa-symmetry gauge \ci{Drukker:2000ep}. 

\def \no {\nonumber}
 Following \cite{Forini:2015bgo,Faraggi:2016ekd,Cagnazzo:2017sny} we get rid of the conformal factor in the induced metric (\ref{g-induced}) by a Weyl transformation. Because the conformal factors at hand (\ref{confactors}) diverge at the symmetry point of the minimal surface, this transformation is actually singular and changes the topology of the worldsheet from the disk to a semi-infinite cylinder. Singularity in the Weyl transformation induces an IR anomaly which affects the one-loop corrections and has to be carefully taken into account \cite{Cagnazzo:2017sny}.
 
 Assuming the standard dependence of the fluctuation operators on the conformal factor,
\begin{equation}\label{conformal}
 \mathcal{K}=\frac{1}{\Omega ^2}\,\widetilde{\mathcal{K}},
 \qquad \ \ \ 
  \mathcal{D}=\frac{1}{\Omega ^{\frac{3}{2}}}\widetilde{\mathcal{D}}\Omega ^{\frac{1}{2}},
\end{equation}
the flat-metric expressions\footnote{We adopt notations and conventions of \cite{Forini:2015bgo}.} for the circle take the form:   
\begin{eqnarray}\label{circleK}
 \CC: \ \ \ \ \  \qquad \qquad \widetilde{\mathcal{K}}_1&=&-\partial _\tau ^2-\partial _\sigma ^2+\frac{2}{\sinh^2\sigma },
\nonumber \\
 \widetilde{\mathcal{K}}_2=\widetilde{\mathcal{K}}_{3\pm}&=&-\partial _\tau ^2-\partial _\sigma ^{2},
\nonumber \\
 \widetilde{\mathcal D}_ \pm  &=&- i{\partial _\tau }{\tau _2} + i{\partial _\sigma }{\tau _1} + \frac{1}{{\sinh \sigma }}{\tau _3}\ , 
\end{eqnarray}
 where $\tau _i$ are the standard Pauli matrices. For the latitude we have:
\begin{eqnarray}\label{latitudeK}
\LL:\qquad  \widetilde{\mathcal{K}}_1&=&  - \partial _\tau ^2 - \partial _\sigma ^2 + \frac{2}{{{{\sinh }^2}\sigma }}, \quad \quad \ \quad \qquad \qquad \qquad  \quad \quad 
\nonumber \\
\widetilde {\mathcal{K}}_{2}&=&  - \partial _\tau ^2 - \partial _\sigma ^2 - \frac{2}{{{{\cosh }^2}\sigma }},
 \\
\widetilde{\mathcal{K}}_{3\pm} &= & - \partial _\tau ^2 - \partial _\sigma ^2 \pm 2i\left( {\tanh  {2\sigma } - 1} \right){\partial _\tau } + \left( {\tanh{2\sigma }- 1} \right)\left( {1 + 3\tanh{2\sigma }} \right)
\vphantom{\frac{2}{{{{\cosh }^2}\sigma }}},
\nonumber \\
 \widetilde{\mathcal D}_ \pm  &=&
  - 
 \left[ {i{\partial _\tau } \mp \frac{1}{2}\left( {1 - \tanh 2\sigma } \right)} \right]{\tau _2} 
 + i{\partial _\sigma }{\tau _1}
 + \frac{1}{{{\Omega }\,{{\sinh }^2}\sigma }}{\tau _3} \mp \frac{1}{{{\Omega }\,{{\cosh }^2}\sigma }}.\no 
\end{eqnarray}

The operator $\mathcal{K}_2$, which describes the three transverse fluctuations  on the sphere, has a zero mode associated with the three-parametric degeneracy of the classical string solution.\foot{
For the  general case of the  latitude WL  with $\theta_{0} \neq {\pi \ov 2} $ (cf. \rf{4}) the corresponding
$\mathcal{K}_{2}$ operator has a zero mode corresponding to
 $\tanh(\sigma+\sigma_{0})$  where $\s_0$ is defined by  $\tanh \sigma_0 =\cos \theta_0$.
  This mode  corresponding  also to the $p\rightarrow
 0$ limit of the continuum spectrum, is not, however,  normalizable.}
 It is easy to see directly that $\psi _0$ defined in (\ref{psi-noll}) satisfies
\begin{equation}
 \mathcal{K}_2\,  
 \psi _0 =0 .
\end{equation}
As  was shown in \cite{Drukker:2000ep} and explicitly checked in \cite{Cagnazzo:2017sny}, the one-loop partition function for the circle,  $\mathop{\mathrm{Sdet}}{}^{-\frac{1}{2}}\mathbbm{K}_{\rm C}$, is Weyl-invariant. For the latitude the situation is more complicated because $\mathcal{K}_2$ has a zero mode. In appendix~\ref{ConfAnomaly} we show that the ratio $\left\langle \psi _0\left|\Omega ^2\right|\psi _0\right\rangle/\det\nolimits'{}\mathcal{K}_2$ shifts under Weyl rescalings by the conventional conformal anomaly, which is eventually canceled by the contributions of other string modes when they are combined together. 
From this we conclude that for the latitude  $\left\langle \psi _0\left|\Omega _L^2\right|\psi _0\right\rangle^{3}/\mathop{\mathrm{Sdet}}\nolimits'{}\mathbbm{K}_\LL$ is Weyl invariant.

Conformal anomaly cancellation allows us to drop the conformal factor in the ratio (\ref{WL/WC})  and to deal with the operators (\ref{circleK}), (\ref{latitudeK}) defined on a flat half-cylinder:
\begin{equation}\label{tilded-WL/WC}
 \frac{W_\LL}{W_\CC}=\frac{1}{4\pi }\,\lambda ^{\frac{3}{4}}
 \,{\rm e}\,^{-\sqrt{\lambda }}\left\langle \psi _0\right.\!\!\left| \psi _0\right\rangle^{\frac{3}{2}}\,\frac{\mathop{\mathrm{Sdet}}{}^\frac{1}{2}\widetilde{\mathbbm{K}}_\CC}{\mathop{\mathrm{Sdet}}'{}^\frac{1}{2}\widetilde{\mathbbm{K}}_\LL}.
\end{equation}
This expression, taken at face value, is ill-defined due to IR divergences at large values of $\sigma $. A way to regularize these divergences is to artificially impose  some  boundary  conditions (e.g.  Dirichlet or Neumann) 
at  a large $\sigma =R_{\CC,\LL}$ and send $R_{\CC,\LL}\rightarrow \infty$ at the end of the calculation. As shown in \cite{Cagnazzo:2017sny}, such regularization does not 
spoil conformal anomaly cancellation. While the cutoff dependence eventually cancels in the ratio (\ref{tilded-WL/WC}), the diffeomorphism invariance requires to choose different cutoffs for the circle and the latitude. The difference leaves a finite remnant in the ratio of
partition functions \cite{Cagnazzo:2017sny}. 

An invariant cutoff is the worldsheet area left out by regularization:
\begin{equation}
 s=\int_{\sigma >R}^{}d^2\sigma \,\sqrt{h}=2\pi \int_{R}^{\infty }
 d\sigma \,\Omega ^2,
\end{equation}
and this should be the same for the circle and the latitude.
Using explicit form (\ref{confactors}) of the scale factors for the respective minimal surfaces, we find that the coordinate cutoffs are related by a finite relative shift:
\begin{equation}\label{RC-RL}
 R_{\rm C}=R_{\rm L}-\frac{1}{2}\,\ln 2.
\end{equation}
Later we will give another justification for this relationship.

The IR cutoff is necessary to define determinants on the half-cylinder. In addition, it regularizes the zero mode normalization
(cf. \rf{psi-noll},\rf{curved<>})
\begin{equation}\label{zero-norm}
 \left\langle \psi _0\right.\!\!\left| \psi _0\right\rangle=2\pi \int_{0}^{R}
 d\sigma \,\tanh^2\sigma =2\pi R+\mathcal{O}(1).
\end{equation}

\section{Determinants and phase-shifts}

The determinants in (\ref{tilded-WL/WC}) were calculated in \cite{Forini:2015bgo,Faraggi:2016ekd,Cagnazzo:2017sny} for a more general, 
non-supersymmetric (but superconformal)  $\theta_0$-latitude  minimal surface  and the results 
   relevant for our problem may  be read off from there by taking the limit  $\theta_0={\pi\ov2}$. More precisely, this works
   directly  for all  the operators except $\mathcal{K}_2$ which develops a zero mode in the supersymmetric  limit. 
   The determinant of $\widetilde{\mathcal{K}}_2$, as a result, blows up in the 
    limit and has to be reconsidered anew by carefully handling the zero mode. 

\subsection{Spectral problem for $\widetilde{\mathcal{K}}_2$}

The eigenvalues of $\widetilde{\mathcal{K}}_2$ defined in (\ref{circleK}),(\ref{latitudeK}) are of the form $\omega ^2+p^2$ where $\omega $ is an integer ``Matsubara"  frequency (corresponding to the  $\tau$-circle) 
and $p^2$ is an eigenvalue of a one-dimensional Schr\"odinger operator defined on  functions of 
 $\sigma$. For the circle this operator is  simply  $- \partial^2_\sigma$, while for the latitude the Schr\"odinger equation is
\begin{equation}\label{Schop}
 \left(-\partial _\sigma ^2-\frac{2}{\cosh^2\sigma }\right)\psi =p^2\psi .
\end{equation}
The solution satisfying the boundary condition $\psi (0)=0$ is
\begin{equation}\label{wavey}
 \psi (\sigma )=\ p\,\sin p\sigma +\tanh\sigma \,\cos p\sigma .
\end{equation}
Notice that for $p=0$ we recover the zero mode eigenfunction $\psi _0$ in \rf{psi-noll}.

The operator (\ref{Schop}) has a continuous spectrum in infinite volume. To define the determinant we need to regularize the problem which we do by imposing the Neumann boundary condition at $\sigma =R_\LL$:
\begin{equation}\label{Neumann}
 {\psi }'(R_\LL)=0.
\end{equation}
The precise form of the boundary condition is not so important, but it has to be consistent with the existence of the zero mode. The Dirichlet condition $\psi (R_\LL)=0$, for instance, eliminates the zero mode and, if imposed, spoils connection to the original spectral problem for the undeformed operator $\mathcal{K}_2$.

A more rigorous regularization prescription consists in gradually eliminating the scale factor by replacing $\Omega $ with $\Omega ^\alpha $  in (\ref{conformal}), as done in the appendix~\ref{ConfAnomaly}. Weyl invariance of the string partition function guarantees that the answer does not depend on $\alpha $, and one can eliminate the scale factor by taking $\alpha \rightarrow 0$. The scale factor for a very small but finite $\alpha $ is approximately equal to one on a wide interval $\,{\rm e}\,^{-1/2\alpha }\ll \sigma \ll 1/\alpha $. The divergence of $\Omega ^\alpha $ at small sigma reinforces the boundary condition $\psi (0)=0$, while exponential decay  at  $\sigma \sim 1/\alpha $ introduces an effective IR cutoff. Defining the cutoff scale through $\Omega ^\alpha \simeq C$, we find that $R\simeq \ln C/\alpha +\nu/2 $, where $\nu =2\ln 2$ for the circle and $\nu =3\ln 2$ for the latitude, giving an alternative explanation to the relationship (\ref{RC-RL}) between the respective IR cutoffs. With some effort these arguments can be made more precise; here  we will not go deeper into the technical details, which are relatively sophisticated,  but  at the end give the same result as the simple-minded Neumann boundary condition.

At large $\sigma $ the wavefunction (\ref{wavey}) assumes  the asymptotic form, valid up to exponential corrections:
\begin{equation}
 \psi (\sigma )\simeq \sqrt{p^2+1}\,\sin\big(p\sigma +\delta (p)\big),
\end{equation}
with the phase-shift equal to
\begin{equation}
 \delta (p) =\frac{\pi }{2}-\arctan p.
\end{equation}
The Neumann boundary condition imposes momentum quantization
\begin{equation}\label{mom-q}
 p_nR_\LL+\delta (p_n)=\pi \Big(n+\frac{1}{2}\Big).
\end{equation}
In this language the zero mode ($p_0=0$)  arises because $\delta (0)={\pi \ov 2}$.

The determinant of $\widetilde{\mathcal{K}}_2$ requires also a UV regularization at large momenta. We will use Pauli-Villars regularization, following \cite{tHooft:1976snw}, and define the regularized determinant by
\begin{equation}
 \det\nolimits_{\rm reg}\widetilde{\mathcal{K}}_2^\LL=\frac{\det\nolimits'{}\widetilde{\mathcal{K}}_2^\LL}{\det\left(\widetilde{\mathcal{K}}_2^\LL+M^2\right)}=
 \frac{1}{M^2}\,\prod_{(\omega ,n)\neq (0,0)}^{}
 \frac{\omega ^2+p_n^2}{\omega ^2+p_n^2+M^2}\,,
\end{equation}
which can also be written as
\begin{equation}\label{detreg}
 \det\nolimits_{\rm reg}\widetilde{\mathcal{K}}_2^\LL=\frac{1}{M^2}\prod_{\omega \neq 0}^{}\prod_{n=0}^{\infty }\frac{\omega ^2+p_n^2}{\omega ^2+p_n^2+M^2}
 \,\,\prod_{n=0}^{\infty }\frac{p_{n+1}^2}{p_{n+1}^2+M^2}\,.
\end{equation}
Eventually we will divide this by the corresponding determinant for the circle, the wavefunction for which
is $\psi (\sigma )=\sin p\sigma $ and the momentum quantization condition
is just
\begin{equation}
 \bar{p}_nR_\CC=\pi \Big(n+\frac{1}{2}\Big).
\end{equation}
 This has no zero mode, and the term with $(\omega ,n)=(0,0)$ need not be excluded from the product:
\begin{equation}
 \det\nolimits_{\rm reg}\widetilde{\mathcal{K}}_2^\CC= \prod_{\omega ,n}^{}
 \frac{\omega ^2+\bar{p}_n^2}{\omega ^2+\bar{p}_n^2+M^2}\,.
\end{equation}

The difference between $R_\LL$ and $R_\CC$, given by (\ref{RC-RL}), can be absorbed into  the redefinition of the phase-shift:
\begin{equation}
 \delta (p)\rightarrow \delta (p)+\frac{p}{2}\,\ln 2,
\end{equation}
after which $R_\CC$ can be used as an IR cutoff for both the circle and the latitude;  we  will denote  it 
simply by $R$ to simplify the notations. The momentum quantization condition (\ref{mom-q}) then reads:
\begin{equation}\label{delta-p}
 p_n-\bar{p}_n=-\frac{1}{R}\left(\delta (p_n)+\frac{p_n}{2}\,\ln 2\right).
\end{equation}

The product in (\ref{detreg}) receives contributions from the  two type of modes, the regularized continuous spectrum with $n\sim R$ and the 
near-zero modes with $n\ll R$. To separate the two we introduce an intermediate scale $N\gg 1$ such that
\begin{equation}\label{eps-def}
 \varepsilon =\frac{\pi N}{R}\ll 1,
\end{equation}
and treat separately the modes with $n<N$ and $n\geq N$.
The near-zero modes are only important at zero Matsubara frequency $\omega=0$. Their contribution to the ratio of determinants is
\begin{equation}\label{low-mom}
  \prod_{n=0}^{N-1}\frac{p^2_{n+1}}{\bar{p}_n^2}=\prod_{n=0}^{N}\frac{\left(n+1\right)^2}{(n+\frac{1}{2})^2}
 =\frac{\Gamma ^2\left(\frac{1}{2}\right)\Gamma ^2\left(N+1\right)}{\Gamma ^2\left(N+\frac{1}{2}\right)}\simeq \pi N.
\end{equation}
The meaning of this formula is simple. For the circle, the wavefunction is $\sin p\sigma $, and the Neumann boundary condition picks up 
 modes with $\bar p_n=\pi (n+{1\ov2})/R$, \  $n=0,1,2,\ldots $  The first term in the latitude wavefunction (\ref{wavey}) is suppressed at low momenta, while the second term asymptotes to $\cos p\sigma $, which gives $p_n=\pi n/R$, $n=0,1,2,\ldots $. The difference in the quantization conditions can be attributed to the non-vanishing phase-shift at zero momentum, for the latitude: $\delta (0)={\pi \ov 2}$. The zero mode for the latitude should be dropped from the product which amounts to shifting the mode number: $n\rightarrow n+1$. This shift affects only the zero Matsubara frequency and can be equivalently  described as a phase-shift redefinition:
\begin{equation}\label{shiftpi}
 \delta (p)\rightarrow \delta (p)-\pi .
\end{equation}

For the contribution of the continuous spectrum, the summation over $n$ can be replaced by integration, and with the help of (\ref{delta-p}) we get:
\begin{equation}
 \sum_{n=N}^{\infty }\big(f(p_n)-f(\bar{p}_n)\big)=
 -\int_{\varepsilon }^{\infty }\frac{dp}{\pi }\,\, 
 {f}'(p)\Big(\delta (p)+\frac{p}{2}\,\ln 2\Big),
\end{equation}
up to corrections that vanish at $R\rightarrow \infty $. Applying this formula to (\ref{detreg}), taking into account an extra shift (\ref{shiftpi}) for the zero Matsubara frequency and adding the low-momentum contribution (\ref{low-mom}), we get for the ratio of determinants:
\begin{eqnarray}
 \frac{\det_{\rm reg}\widetilde{\mathcal{K}}_2^\LL}{\det_{\rm reg}\widetilde{\mathcal{K}}_2^\CC}
 &=&\frac{\pi N}{M^2}\exp\Big[
 - 2\int_{\varepsilon }^{\infty }\frac{dp}{\pi }\Big(
 \delta (p)+\frac{p}{2}\,\ln 2\Big)
 \sum_{\omega \neq 0}^{}\Big(\frac{p}{\omega ^2+p^2}-\frac{p}{\omega ^2+p^2+M^2}\Big) 
\nonumber \\
&& 
 -\, 2\int_{\varepsilon }^{\infty }\frac{dp}{\pi }\,
 \Big(\delta (p)+\frac{p}{2}\,\ln 2-\pi \Big)
 \Big(\frac{1}{p}-\frac{p}{p^2+M^2}\Big)
  \Big].
\end{eqnarray}

As a next step, we perform  summation over $\omega $ and explicitly integrate  the $\pi $ term in the last integral. 
The UV logarithm from this integral neatly cancels the UV divergence due to the zero mode ($1/M^2$ in the prefactor),  which is not  surprising
as they both originate from the ``spectral flow": omitted zero mode leaves an uncompensated mode in the regulator determinant and, at the same time, shifts up mode numbers by one, which is ultimately responsible for the extra UV log.
The IR part of the log combines with the auxiliary scale $N$ to form the physical IR cutoff $R$.  Using the definition (\ref{eps-def}), we find:
\begin{eqnarray}
 \frac{\det_{\rm reg}\widetilde{\mathcal{K}}_2^\LL}{\det_{\rm reg}\widetilde{\mathcal{K}}_2^\CC}
 &=&R\,\exp\Big[
 -2\int_{\varepsilon }^{\infty }dp\,\Big(\delta (p)+\frac{p}{2}\,\ln 2\Big)
\nonumber \\
&&\times 
\ \Big(\coth\pi p-\frac{p}{\sqrt{p^2+M^2}}\,\coth\pi \sqrt{p^2+M^2}\Big)
\vphantom{2\int_{\varepsilon }^{\infty }dp\,\Big(\delta (p)+\frac{p}{2}\,\ln 2 \Big)
 \Big(\coth\pi p-\frac{p}{\sqrt{p^2+M^2}}\coth\pi \sqrt{p^2+M^2}\Big)}
-\ln\varepsilon 
 \Big].
\end{eqnarray}
The dependence on $\varepsilon$ in the exponent is fake, it actually cancels out because the momentum integral is log divergent on the lower limit. 

\subsection{Phase-shifts and other operators}

The determinants of  the other operators  in  (\ref{latitudeK}) are substantially simpler, because they have no zero modes, and can be expressed through the phase-shifts for the respective 1d operators. 
The explicit expressions for the phase-shifts (taken from \cite{Cagnazzo:2017sny}) are:
\begin{eqnarray}
\CC: \qquad \qquad   \delta _1(p)&=&-\arctan p\ , 
\nonumber \\
 \delta _2(p)&=&0\ , 
\nonumber \\
\delta _3(p)&=&0\ , 
\nonumber \\
 \delta _F(p)&=&-\arctan 2p\ ,  
\end{eqnarray}
for the circle and
\begin{eqnarray}
\LL: \qquad \qquad   \delta _1(p)&=&-\arctan p\ , 
\nonumber \\
 \delta _2(p)&=&\frac{\pi }{2}-\arctan p\ , 
\nonumber \\
 \delta_3(p)&=&\frac{1}{2}\,\arctan p-\frac{1}{2}\,\arctan\frac{p}{2}\ , 
\nonumber \\
\delta _F(p)&=&-\frac{1}{2}\,\arctan 2p-\frac{1}{2}\,\arctan\frac{2p}{3} \ , 
\end{eqnarray}
for the latitude. The only mode with $\delta (0)\neq 0$ is the one corresponding to  $\widetilde{\mathcal{K}}_2$, all
 others have $\delta (0)=0$ and do not produce an IR contribution similar to (\ref{low-mom}).

Combining the  contributions of all the modes together we find:
\begin{eqnarray}
 \frac{\mathop{\mathrm{Sdet}}'\mathbbm{K}_\LL}{\mathop{\mathrm{Sdet}}\mathbbm{K}_\CC}&=&R^3\,\exp\Big[
 -2\int_{\varepsilon }^{\infty }dp\,\sum_{a}^{}\left(-1\right)^{F_a}
 \Big(\delta _a^\LL(p)-\delta _a^\CC(p)+\frac{p}{2}\,\ln 2\Big)
\nonumber \\
&&\qquad \qquad\qquad \qquad  \times  
 \coth^{\left(-1\right)^{F_a}}\pi p
\vphantom{-2\int_{\varepsilon }^{\infty }dp\,\sum_{a}^{}\left(-1\right)^{F_a}
 \Big(\delta _a^\LL(p)-\delta _a^\CC(p)+\frac{p}{2}\,\ln 2\Big)
 \coth^{\left(-1\right)^{F_a}}\pi p}
-3\ln\varepsilon 
 \Big], 
\end{eqnarray}
where the sum is over all $8_{\rm B}+8_{\rm F}$ string modes counted with multiplicities. This formula takes into account anti-periodic boundary conditions for fermions in $\tau $, which entails half-integer Matsubara frequencies and replacement of $\coth$ by $\tanh$ in the momentum integral. 
The UV divergences in the momentum integral happen to cancel, allowing us to drop the regulator term. The IR scale $R$ cancels in (\ref{tilded-WL/WC}) the divergent norm of the zero-mode wavefunction (\ref{zero-norm})\footnote{The IR cutoff in (\ref{zero-norm}) is $R_\LL$ while the prefactor in the ratio of determinants is $R_\CC$. Their difference, of order one, was important to keep in the exponent, but in the pre-factor this difference is immaterial as long as $R\rightarrow \infty $.}.

The final result for the semiclassical string theory  expression   for the ratio of the two Wilson loops   is thus 
\begin{equation}\label{final-WL/WC}
 \frac{W_\LL}{W_\CC}=\sqrt{\frac{\pi }{2}}\,\lambda ^{\frac{3}{4}}
 \,{\rm e}\,^{-\sqrt{\lambda }}\ Z_{\rm 1-loop}\  ,
\end{equation}
where $Z_{\rm 1-loop}$    given by 
\begin{equation}
 \ln Z_{\rm 1-loop}=\int_{\varepsilon }^{\infty }dp\,\sum_{a}^{}
 \left(-1\right)^{F_a}
 \Big(\delta _a^\LL(p)-\delta _a^\CC(p)+\frac{p}{2}\,\ln 2\Big)
 \coth^{\left(-1\right)^{F_a}}\pi p
 +\frac{3}{2}\,\ln\varepsilon \la{Z}
\end{equation}
is the genuine contribution of the string fluctuations. 
It is interesting to note that 
the zero modes produce the factor  $\sqrt{\frac{\pi }{2}}\,\lambda ^{\frac{3}{4}}$  in (\ref{final-WL/WC}), which  itself happens 
to agree  already with the prefactor  in  the    gauge-theory prediction
in \rf{prediction}.

\subsection{String fluctuations}

To compute the momentum integral in (\ref{Z}) we first see that
\begin{eqnarray}\label{sumofphase-shifts}
 &&\sum_{a}^{}
\left(-1\right)^{F_a}
 \left(\delta _a^\LL(p)-\delta _a^\CC(p)+\frac{p}{2}\,\ln 2\right)
\nonumber \\
 &&=\left(\frac{3\pi }{2}-\arctan\frac{p}{2}-2\arctan p\right)_B
 +4\left(\arctan \frac{2p}{3}-\arctan 2p\right)_F.
\end{eqnarray}
The combined expression decreases very fast at infinity, as $1/p^3$, and can be integrated just by itself:
\begin{equation}
J_{\rm I}= \int_{0}^{\infty }dp\,\left(-1\right)^{F_a}
 \left(\delta _a^\LL(p)-\delta _a^\CC(p)+\frac{p}{2}\,\ln 2\right)
 =6\ln\frac{3}{2} 
\ .\la{I} \end{equation}
In the full integral \rf{Z}, the bosonic/fermionic part of (\ref{sumofphase-shifts}) is multiplied by $\coth\pi p$/$\tanh\pi p$. After subtracting $1$ from the hyperbolic functions,  each term in the sum converges individually and all the terms can be integrated one by one. The basic integrals are
\begin{eqnarray}
J_{\rm II} = \int_{0}^{\infty }dp\,\left(\coth\pi p-1\right)\arctan\frac{p}{z}
 =\ln\Gamma (z)-\Big(z-\frac{1}{2}\Big)\ln z+z-\frac{1}{2}\,\ln 2\pi \ , 
\nonumber \\
J_{\rm III}=\int_{0}^{\infty }dp\,\left(\tanh\pi p-1\right)\arctan\frac{p}{z}
 =\ln\Gamma \Big(z+\frac{1}{2}\Big)-z\ln z+z-\frac{1}{2}\,\ln 2\pi \ . \qquad  
\end{eqnarray}
The constant term produces an IR divergent integral: 
\begin{equation}
J_{\rm IV}= \pi \int_{\varepsilon }^{\infty }dp\,\left(\coth\pi p-1\right)
 =-\ln (2\pi \varepsilon) \ . 
\end{equation}
Finally,  there is also a contribution from the common auxiliary phase-shift originating from the difference \rf{RC-RL}  between $R_\LL$ and $R_\CC$:
\begin{equation}
J_{\rm V}=  8\int_{0}^{\infty }dp\,p\left(\coth\pi p-\tanh\pi p\right)=1 \ . \la{V}
\end{equation}
Collecting all the pieces together we get:
\begin{eqnarray}
 \ln Z_{\rm 1-loop}&=&\Big(6\ln\frac{3}{2}\Big)_{\rm (I)}
 +\Big(\frac{3}{2}\,\ln 2\pi -\frac{1}{2}\,\ln 2-6\ln\frac{3}{2}\Big)_{\rm (II)+(III)}
 \nonumber \\
 &&+\Big(-\frac{3}{2}\,\ln (2\pi \varepsilon) \Big)_{\rm (IV)}
+\Big(\frac{1}{2}\,\ln 2\Big)_{\rm (V)}+\frac{3}{2}\ln \varepsilon ,
\end{eqnarray}
where subscripts indicate  
 individual contributions  of the  integrals \rf{I}--\rf{V}. 
 When the dust settles,  we find
\begin{equation}
 Z_{\rm 1-loop}=1,
\end{equation}
so that  \rf{final-WL/WC}  becomes 
\begin{equation}\label{final-WL/WC2}
 \frac{W_\LL}{W_\CC}=\sqrt{\frac{\pi }{2}}\,\lambda ^{\frac{3}{4}}
 \,{\rm e}\,^{-\sqrt{\lambda }},
\end{equation}
in agreement with the field-theory prediction \rf{prediction}.

\section{Conclusions}

We have shown how string theory in $AdS_5\times S^5$ reproduces the strong-coupling asymptotics of the exact expectation value of the circular Wilson loop, which has been known for a long time \cite{Erickson:2000af,Drukker:2000rr,Pestun:2007rz}, finally nailing down the exact prefactor which arises due to quantum fluctuations of the string. The derivation is not very simple and rests on an additional assumption that the supersymmetric latitude has trivial expectation value. The latter is a theorem in field theory \cite{Guralnik:2003di,Guralnik:2004yc}, but an intrinsic string-theory derivation of this statement is lacking. Perhaps it can be proven by extending the classical argument of \cite{Dymarsky:2006ve}, that applies to general BPS surfaces, to quantum theory.

It is interesting that the whole contribution to the prefactor comes from the zero modes, while the non-zero modes cancel, as they do for the straight line related to the circle by an (anomalous) conformal transformation. The nature of this cancellation may shed light on a non-perturbative disc amplitude, which is supposed to reproduce the whole Bessel function expression (\ref{1}), valid at any coupling in planar super-Yang-Mills theory.

\subsection*{Acknowledgements}
We would like to thank Y.~Makeenko for interesting discussions. The work of D.~M.~R. and K.~Z. was 
supported by the ERC advanced grant No 341222. The work of K.~Z. was in addition supported by the Swedish Research Council (VR) grant
2013-4329, by the grant  ``Exact Results in Gauge and String Theories" from the Knut and Alice Wallenberg foundation, and by RFBR grant 18-01-00460 A. K.Z. was partially supported by the Simons Foundation 
under the program  Targeted Grants to Institutes  (the Hamilton Mathematics 
Institute). K.Z. thanks the Galileo Galilei Institute for Theoretical Physics for 
hospitality during the course of this work and is grateful to INFN and Simons Foundation for partial support.
AAT  was  supported by the STFC grant ST/P000762/1  and also 
 by   the Russian Science Foundation grant 14-42-00047 at Lebedev Institute.

\appendix

\section{Conformal anomaly and zero modes}
\label{ConfAnomaly}

Consider a family of 2d scalar operators ($\phi$ and $E$ are some given functions) 
\begin{equation}
 K(\alpha )=\,{\rm e}\,^{2\alpha\phi }\left(-\partial ^2+E\right),
\end{equation}
each having   some number of 
zero modes:
\begin{equation}
 K(\alpha )\left|n\right\rangle=0.
\end{equation}
The operator $K(\alpha )$ is Hermitian with respect to the measure
\begin{equation}
 \left\langle \psi _1,\psi _2\right\rangle_\alpha =
 \int_{}^{}d^2\sigma \,\,{\rm e}\,^{-2\alpha \phi }\psi _1^*\psi _2.
\end{equation}
As in the main text, we reserve the bra-ket notations for the conventional ($\alpha =0$) scalar product. The projector onto non-zero modes of $K(\alpha )$ then takes the form
\begin{equation}
 P=\mathbbm{1}-\sum_{n}^{}\frac{\left|n\right\rangle\left\langle n\right|\,{\rm e}\,^{-2\alpha \phi }}{\left\langle n\right|\,{\rm e}\,^{-2\alpha \phi }\left|n\right\rangle}\,.
\end{equation}
The dependence of the determinant of $K$ on $\alpha $ is governed by the conformal anomaly. Here we give a brief derivation, emphasizing the contribution of the zero modes, following \cite{Fursaev-Vassilevich}. 

The regularized determinant of $K$ (with zero modes omitted) can be defined via its zeta-function:
\begin{equation}
 \ln\det\nolimits '{}K=-\lim_{s\rightarrow 0}\,\frac{d}{ds}\,\,
 \frac{1}{\Gamma (s)}\int_{0}^{\infty }dt\,t^{s-1}\mathop{\mathrm{Tr}}P\,{\rm e}\,^{-tK}.
\end{equation}
Differentiating in $\alpha $, and taking into account that
\begin{equation}
 \frac{\partial }{\partial \alpha }\,\mathop{\mathrm{Tr}}P\,{\rm e}\,^{-tK}
 =2t\,\frac{\partial }{\partial t}\,\mathop{\mathrm{Tr}}P\phi \,{\rm e}\,^{-tK},
\end{equation}
we get
\begin{eqnarray}
 \frac{d}{d\alpha }\,\ln\det\nolimits '{}K&=&2\lim_{s\rightarrow 0}\frac{d}{ds}
 \frac{s}{\Gamma (s)}\int_{0}^{\infty }dt\,t^{s-1}\mathop{\mathrm{Tr}}P\phi \,{\rm e}\,^{-tK}=2\lim_{t\rightarrow 0}\mathop{\mathrm{Tr}}P\phi \,{\rm e}\,^{-tK}
\nonumber \\
&=&
2\lim_{t\rightarrow 0}\mathop{\mathrm{Tr}} (\phi \,{\rm e}\,^{-tK})
-2\sum_{n}^{}\frac{\left\langle n\right|\phi \,{\rm e}\,^{-2\alpha \phi }\left|n\right\rangle}{\left\langle n\right|\,{\rm e}\,^{-2\alpha \phi }\left|n\right\rangle}\,.
\end{eqnarray}
The first term can be expressed in terms of the second DeWitt-Seeley coefficient of $K$ (see \cite{Fursaev-Vassilevich} for more details), giving
\begin{equation}
 \frac{d}{d\alpha }\,\ln\det\nolimits '{}K=-\frac{1}{2\pi }\int_{}^{}d^2\sigma \,
 \left(\frac{\alpha }{3}\,\partial _\mu \phi \partial ^\mu \phi +\phi E\pm\frac{1 }{2}\,\partial _\mu \partial ^\mu \phi \right)
 -2\sum_{n}^{}\frac{\left\langle n\right|\phi \,{\rm e}\,^{-2\alpha \phi }\left|n\right\rangle}{\left\langle n\right|\,{\rm e}\,^{-2\alpha \phi }\left|n\right\rangle}\,,
\end{equation}
where the $\pm$ sign reflects the difference between the Neumann/Dirichlet boun\-dary conditions at the endpoints of the string.
Integration over $\alpha $ gives:
\begin{equation}
 \ln\frac{\det\nolimits '{}K(1)}{\det\nolimits '{}K(0)}=
 -\frac{1}{2\pi }\int_{}^{}d^2\sigma \,
 \left(\frac{1}{6}\,\partial _\mu \phi \partial ^\mu \phi +\phi E\pm\frac{1 }{2}\,\partial _\mu \partial ^\mu \phi \right)
 +\sum_{n}^{}\ln\frac{\left\langle n\right|\,{\rm e}\,^{-2\phi }\left|n\right\rangle}{\left\langle n\right.\!\!\left| n\right\rangle}\,.
\end{equation}

This result shows that  the ratio
\begin{equation}
 \frac{\det\nolimits '{}K(1)}{\prod\limits_{n}^{}\left\langle n\right|\,{\rm e}\,^{-2\phi }\left|n\right\rangle}
\end{equation}
transforms under Weyl rescalings the same way a determinant without zero modes would transform.

\bibliographystyle{nb}

\begin{thebibliography}{10}
\ifx\href\asklfhas\newcommand{\href}[2]{#2}\fi
\raggedright
\small
\parskip 0pt

\bibitem{Polyakov:1997tj}
A.~M.~Polyakov,
\textit{``{String theory and quark confinement}''},
\textsf{Nucl.~Phys.~Proc.~Suppl.~68,~1~(1998)},
\href{http://arXiv.org/abs/hep-th/9711002}{\texttt{hep-th/9711002}}.
%
\bibitem{Maldacena:1998im}
J.~M.~Maldacena,
\textit{``{Wilson loops in large N field theories}''},
\textsf{Phys.~Rev.~Lett.~80,~4859~(1998)},
\href{http://arXiv.org/abs/hep-th/9803002}{\texttt{hep-th/9803002}}.
%
\bibitem{Rey:1998ik}
S.-J.~Rey and J.-T.~Yee,
\textit{``{Macroscopic strings as heavy quarks in large N gauge theory and
  anti-de Sitter supergravity}''},
\textsf{Eur.~Phys.~J.~C22,~379~(2001)},
\href{http://arXiv.org/abs/hep-th/9803001}{\texttt{hep-th/9803001}}.
%
\bibitem{Alvarez:1982zi}
O.~Alvarez,
\textit{``{Theory of Strings with Boundaries: Fluctuations, Topology, and
  Quantum Geometry}''},
\textsf{Nucl.~Phys.~B216,~125~(1983)}.
%
\bibitem{Berenstein:1998ij}
D.~E.~Berenstein, R.~Corrado, W.~Fischler and J.~M.~Maldacena,
\textit{``{The operator product expansion for Wilson loops and surfaces in the
  large N limit}''},
\textsf{Phys.~Rev.~D59,~105023~(1999)},
\href{http://arXiv.org/abs/hep-th/9809188}{\texttt{hep-th/9809188}}.
%
\bibitem{Drukker:1999zq}
N.~Drukker, D.~J.~Gross and H.~Ooguri,
\textit{``{Wilson loops and minimal surfaces}''},
\textsf{Phys.~Rev.~D60,~125006~(1999)},
\href{http://arXiv.org/abs/hep-th/9904191}{\texttt{hep-th/9904191}}.
%
\bibitem{Erickson:2000af}
J.~K.~Erickson, G.~W.~Semenoff and K.~Zarembo,
\textit{``{Wilson loops in N = 4 supersymmetric Yang-Mills theory}''},
\textsf{Nucl.~Phys.~B582,~155~(2000)},
\href{http://arXiv.org/abs/hep-th/0003055}{\texttt{hep-th/0003055}}.
%
\bibitem{Drukker:2000rr}
N.~Drukker and D.~J.~Gross,
\textit{``{An exact prediction of N = 4 SUSYM theory for string theory}''},
\textsf{J.~Math.~Phys.~42,~2896~(2001)},
\href{http://arXiv.org/abs/hep-th/0010274}{\texttt{hep-th/0010274}}.
%
\bibitem{Pestun:2007rz}
V.~Pestun,
\textit{``{Localization of gauge theory on a four-sphere and supersymmetric
  Wilson loops}''},
\textsf{Commun.Math.Phys.~313,~71~(2012)},
\href{http://arXiv.org/abs/0712.2824}{\texttt{0712.2824}}.
%
\bibitem{Drukker:2000ep}
N.~Drukker, D.~J.~Gross and A.~A.~Tseytlin,
\textit{``{Green-Schwarz string in $AdS_5\times S^5$: Semiclassical partition
  function}''},
\textsf{JHEP~0004,~021~(2000)},
\href{http://arXiv.org/abs/hep-th/0001204}{\texttt{hep-th/0001204}}.
%
\bibitem{Kruczenski:2008zk}
M.~Kruczenski and A.~Tirziu,
\textit{``{Matching the circular Wilson loop with dual open string solution at
  1-loop in strong coupling}''},
\textsf{JHEP~0805,~064~(2008)},
\href{http://arXiv.org/abs/0803.0315}{\texttt{0803.0315}}.
\bibitem{Kristjansen:2012nz}
C.~Kristjansen and Y.~Makeenko,
\textit{``{More about One-Loop Effective Action of Open Superstring in
  $AdS_5\times S^5$}''},
\textsf{JHEP~1209,~053~(2012)},
\href{http://arXiv.org/abs/1206.5660}{\texttt{1206.5660}}.
%
\bibitem{Buchbinder:2014nia}
E.~Buchbinder and A.~Tseytlin,
\textit{``{The 1/N correction in the D3-brane description of circular Wilson
  loop at strong coupling}''},
\textsf{Phys.Rev.~D89,~126008~(2014)},
\href{http://arXiv.org/abs/1404.4952}{\texttt{1404.4952}}.
%
\bibitem{Ambjorn:2011wz}
J.~Ambjorn and Y.~Makeenko,
\textit{``{Remarks on Holographic Wilson Loops and the Schwinger Effect}''},
\textsf{Phys.~Rev.~D85,~061901~(2012)},
\href{http://arXiv.org/abs/1112.5606}{\texttt{1112.5606}}.
%
\bibitem{Zarembo:2002an}
K.~Zarembo,
\textit{``{Supersymmetric Wilson loops}''},
\textsf{Nucl.~Phys.~B643,~157~(2002)},
\href{http://arXiv.org/abs/hep-th/0205160}{\texttt{hep-th/0205160}}.
%
\bibitem{Guralnik:2003di}
Z.~Guralnik and B.~Kulik,
\textit{``{Properties of chiral Wilson loops}''},
\textsf{JHEP~0401,~065~(2004)},
\href{http://arXiv.org/abs/hep-th/0309118}{\texttt{hep-th/0309118}}.
%
\bibitem{Guralnik:2004yc}
Z.~Guralnik, S.~Kovacs and B.~Kulik,
\textit{``{Less is more: Non-renormalization theorems from lower dimensional
  superspace}''},
\textsf{Int.~J.~Mod.~Phys.~A20,~4546~(2005)},
\href{http://arXiv.org/abs/hep-th/0409091}{\texttt{hep-th/0409091}}.
%
\bibitem{Drukker:2005cu}
N.~Drukker and B.~Fiol,
\textit{``{On the integrability of Wilson loops in $AdS_5\times S^5$: Some
  periodic ansatze}''},
\textsf{JHEP~0601,~056~(2006)},
\href{http://arXiv.org/abs/hep-th/0506058}{\texttt{hep-th/0506058}}.
%
\bibitem{Drukker:2006ga}
N.~Drukker,
\textit{``{1/4 BPS circular loops, unstable world-sheet instantons and the
  matrix model}''},
\textsf{JHEP~0609,~004~(2006)},
\href{http://arXiv.org/abs/hep-th/0605151}{\texttt{hep-th/0605151}}.
%
\bibitem{Pestun:2009nn}
V.~Pestun,
\textit{``{Localization of the four-dimensional N=4 SYM to a two-sphere and 1/8
  BPS Wilson loops}''},
\textsf{JHEP~1212,~067~(2012)},
\href{http://arXiv.org/abs/0906.0638}{\texttt{0906.0638}}.
%
\bibitem{Forini:2015bgo}
V.~Forini, V.~Giangreco, M.~Puletti, L.~Griguolo, D.~Seminara and E.~Vescovi,
\textit{``{Precision calculation of 1/4-BPS Wilson loops in AdS$_5\times
  S^5$}''},
\textsf{JHEP~1602,~105~(2016)},
\href{http://arXiv.org/abs/1512.00841}{\texttt{1512.00841}}.
%
\bibitem{Faraggi:2016ekd}
A.~Faraggi, L.~A.~Pando~Zayas, G.~A.~Silva and D.~Trancanelli,
\textit{``{Toward precision holography with supersymmetric Wilson loops}''},
\textsf{JHEP~1604,~053~(2016)},
\href{http://arXiv.org/abs/1601.04708}{\texttt{1601.04708}}.
%
\bibitem{Forini:2017whz}
V.~Forini, A.~A.~Tseytlin and E.~Vescovi,
\textit{``{Perturbative computation of string one-loop corrections to Wilson
  loop minimal surfaces in AdS$_5 \times$ S$^5$}''},
\textsf{JHEP~1703,~003~(2017)},
\href{http://arXiv.org/abs/1702.02164}{\texttt{1702.02164}}.
%
\bibitem{Cagnazzo:2017sny}
A.~Cagnazzo, D.~Medina-Rincon and K.~Zarembo,
\textit{``{String corrections to circular Wilson loop and anomalies}''},
\textsf{JHEP~1802,~120~(2018)},
\href{http://arXiv.org/abs/1712.07730}{\texttt{1712.07730}}.
%
\bibitem{Dymarsky:2006ve}
A.~Dymarsky, S.~S.~Gubser, Z.~Guralnik and J.~M.~Maldacena,
\textit{``{Calibrated surfaces and supersymmetric Wilson loops}''},
\textsf{JHEP~0609,~057~(2006)},
\href{http://arXiv.org/abs/hep-th/0604058}{\texttt{hep-th/0604058}}.
%
\bibitem{Forini:2015mca}
V.~Forini, V.~G.~M.~Puletti, L.~Griguolo, D.~Seminara and E.~Vescovi,
\textit{``{Remarks on the geometrical properties of semiclassically quantized
  strings}''},
\textsf{J.~Phys.~A48,~475401~(2015)},
\href{http://arXiv.org/abs/1507.01883}{\texttt{1507.01883}}.
%
\bibitem{tHooft:1976snw}
G.~'t~Hooft,
\textit{``{Computation of the Quantum Effects Due to a Four-Dimensional
  Pseudoparticle}''},
\textsf{Phys.~Rev.~D14,~3432~(1976)}.
%
\bibitem{Fursaev-Vassilevich}
D.~Fursaev and D.~Vassilevich,
\textit{``Operators, geometry and quanta''},
Springer (2011).
%
\end{thebibliography}

\end{document}